\documentclass[12pt]{iopart}
\usepackage{amssymb}
\usepackage{iopams,bm,verbatim}
\usepackage[T1]{fontenc}
\usepackage{subeqn}
\usepackage{amsthm}

\def\Journal#1#2#3#4{{#4} {\it #1} {\bf #2}, #3 }

\def\Xc{\bar{X}}
\def\td{\tau'}
\def\tdbar{\overline{\tau}'}
\def\tbar{\overline{\tau}}
\def\rd{\rho'}

\def\phibar{\overline{\phi}}
\newcommand{\w}[1]{\bm{#1}} 

\def\tho{\textrm{\TH}}
\def\thd{\tho '}
\def\et{\eth}
\def\etd{\eth '}
\def\ud{\textrm{d}}
\def\zc{\overline{\zeta}}
\newcommand{\be}{\begin{equation}}
\newcommand{\ee}{\end{equation}}

\begin{document}
%
\title{Homogeneous Killing spinor space-times}

\author{N.~Van den Bergh}

\address{Ghent University, Department of Mathematical Analysis IW16, \\ Galglaan 2, 9000 Ghent, Belgium}
\eads{\mailto{norbert.vandenbergh@ugent.be}}

\begin{abstract}
      A classification of Petrov type D Killing spinor space-times admitting a homogeneous conformal representant is presented. For each class a canonical line-element is constructed 
and a physical interpretation of its conformal members is discussed.

\end{abstract}

\pacs{04.20.Jb}

\section{Introduction}
In the area of exact solutions of the Einstein field equations space-times admitting Killing spinors occupy a special position. Many known and important exact solutions admit Killing 
spinors: the Friedmann-Lema\^\i tre-Robertson-Walker, Kerr, Kantowski-Sachs, Schwarzschild interior and Wahlquist's rotating perfect fluid metrics are just a few examples crossing one's mind. 
More generally every conformally flat space-time admits Killing spinors, as well as every locally rotationally symmetric perfect fluid, every Petrov type D vacuum solution 
and, with the exception of the Plebanski-Hacyan metrics, every Petrov type D (doubly aligned) Einstein-Maxwell solution. As the existence of a Killing spinor is a conformally invariant property
of a space-time, this suggests the construction of new, physically relevant, exact solutions (for example
rotating perfect fluids or non-aligned Petrov type D Einstein-Maxwell solutions), by finding suitable conformal representants of Killing spinor space-times. Because of the particular 
relevance of the Petrov type D situation, a chief concern ---but an undertaking which has not yet been completed--- is the construction of all so-called KS space-times. 
These were
defined in~\cite{McLenVdB1993} as non-conformally flat space-times admitting a \emph{non-null} valence two 
Killing spinor $X_{AB}$. Alternatively they can be characterised as the class of Petrov type D conformal Killing-Yano space-times: their repeated 
principal Weyl spinors are aligned with the principal spinors of $X_{AB}$ \cite{PlebHacyan76},
 \begin{equation} \label{eq1}
 X_{AB}=X o_{(A} \iota_{B)},
 \end{equation}
 and define geodesic shear-free null congruences, while the square $P_{ab}={D_a}^c D_{cb}$ of the conformal Killing-Yano
two-form $D_{ab}= X_{A B} \epsilon_{A'B'}+\Xc_{A'B'} \epsilon_{AB}$ is a conformal Killing tensor of Segre type $[(11)(11)]$. 
There is
always a conformal representant in which $P_{ab}$ is a Killing tensor, namely when the modulus $|X|$ of the Killing spinor is constant 
(by means of a global rescaling this constant can be taken $=1$): this conformal representant will be called the unitary representant. 
To facilitate the discussion a Killing tensor of Segre type $[(11)(11)]$ with two non-constant (double) eigenvalues
will be called regular, in contrast to the ones with one or two constant eigenvalues, which will be called semi-regular 
or exceptional respectively. The Killing tensor existing in the unitary representant belongs to the latter family and vice-versa:
if a representant admits an exceptional Killing tensor, then that representant is --modulo a constant rescaling-- the unitary one.
\emph{Generically} a KS space-time admits conformal representants with a regular Killing tensor: 
these are the space-times considered by Jeffryes~\cite{Jeffryes1984} and which play an important role in for example Einstein-Maxwell theory (for example
 all fully aligned Petrov type D Einstein-Maxwell solutions~\cite{DebeverMcLen1981}, with 
the exception of the Plebanski-Hacyan metrics~\cite{PlebHacyan76}, belong to this category). 
KS space-times which do not possess a conformal representant with a regular Killing tensor necessarily belong to 
Jeffryes'~\cite{Jeffryes1984} classes $I$ 
or $I_N$:
while the semi-regular case was dealt with in \cite{McLenVdB1993} for class $I_N$ and in \cite{KS1_2} for class $I$, the existence of
KS space-times of Jeffryes' class $I$ admitting neither regular nor semi-regular Killing tensors, was demonstrated only
recently~\cite{Beke_et_al2011}. 
These space-times were obtained by imposing some algebraic restrictions on the curvature  
and turned out to be homogeneous, having a 4-dimensional maximal isometry group. 
These results also indicated that a set of space-times might have been overlooked in \cite{McLenVdB1993}: the form of the homogeneous class I metrics suggests the existence
of a homogeneous limit of class $I_N$, contradicting the property that all $I_N$ space-times of \cite{McLenVdB1993} admit a 3-dimensional  maximal isometry group. Indeed, it turned 
out that in \cite{McLenVdB1993} the possibility was 
overlooked that all Cartan scalars could be constants, thereby giving rise to a homogeneous space-time. Together with the fact that it was 
not at all obvious from \cite{Jeffryes1984} which of the KS space-times of classes $II, III, III_N,IV$ did admit an isometry group of dimension $\geqslant 4$, this 
seemed to justify a systematic investigation of all KS space-times admitting a homogeneous conformal representant.  Provided that the space-time is of \emph{Petrov type D} (because
 of the resulting alignment of the Killing spinor with the Weyl spinor, see equation (\ref{eq1})), the modulus $|X|$ of the Killing spinor is then a 
geometric invariant and is therefore constant: the homogeneous members of this family are then precisely the unitary 
representant and its constant rescalings. 

In \S 2 I present the main equations describing KS space-times with a homogeneous unitary representant. 
In \S 3  I give a complete classification of these space-times, 
explicitly listing the corresponding line-elements. In \S 4 a possible physical interpretation of the conformal representants of the 
different classes is discussed.

\section{Main equations}

Following the notations and conventions of \cite{Beke_et_al2011}, the tetrad basis vectors are taken as $\w{k},\w{\ell},\w{m},
\overline{\w{m}}$ with $- k^a \ell_a = 1 = m^a
\overline{m}_a$. The correspondence with the Newman-Penrose operators\cite{NP} and the basis one-forms
is given by $(m^a,\overline{m}^a, \ell^a, k^a) \leftrightarrow (\delta, \overline{\delta}, \Delta, D)$ and
$(\overline{m}_a,m_a,-k_a,-l_a) \leftrightarrow
(\w{\omega}^1,\w{\omega}^2,\w{\omega}^3,\w{\omega}^4)$. 
The description of the problem becomes most compact when using the Geroch-Held-Penrose formalism\footnote{for details the reader is 
referred to \cite{GHP}, but to ease comparison with 
the Newman-Penrose formalism, remember that the GHP weighted operators $\tho,\thd, \et, \etd$ generalise the NP operators 
$D,\Delta, \delta, \overline{\delta}$, while the GHP variables $\kappa',\sigma',\rho', \tau'$ replace the NP variables 
$-\nu,-\lambda, -\mu, -\pi$}: 
writing a symmetric (non-null) spinor $X_{AB}$ as in (\ref{eq1}) and using $o, \iota$ as the 
basis spinors, the property that $X_{AB}$ is a Killing spinor implies (see \cite{Jeffryes1984} for details)
\begin{equation}\label{eqKSa}
\kappa=\sigma=0
\end{equation}
and
\begin{eqnarray}
\tho X = -\rho X, \label{KSb1}\\
\et X = -\tau X, \label{KSb2}
\end{eqnarray}
together with their `primed versions' and $X'=X$, namely $\kappa'=\sigma'=0$ and $\thd X =-\rd X$, $\etd X = -\td X$.
It follows that the Weyl tensor is of Petrov type D (or O) and that $\w{k},\w{\ell},\w{m},\overline{\w{m}}$ are its principal null directions.  

In the unitary representant $|X|$ is constant and hence
\begin{equation}\label{unitary}
 \rho+\overline{\rho} = \tau+\tdbar =0.
\end{equation}
The main equations become then\\

\noindent
a) the integrability conditions expressing the existence of the Killing spinor:
\be \label{specKS}
\thd \rho -\tho \rd =0 , \ \et \td -\etd \tau =0, \ \tho \td -\etd \rho = 0,
\ee
b) the GHP equations:
\begin{eqnarray}
\tho \rho =0, \label{thrho}\\
\et \rho = 2 \rho \tau +\Phi_{01},\label{etrho}
\end{eqnarray}
\begin{eqnarray}
\tho \tau = 2 \rho \tau + \Phi_{01}, \label{thtau} \\
\et \tau = 0, \label{ettau}
\end{eqnarray}
\be
\tho \rd -\et \td
= -\rho \rd -\tau \tbar-\Psi_2-\frac{1}{12} R \label{thrd_ettd}\\
\ee
\begin{eqnarray}
\label{defPhi00}\Phi_{00}=-\rho^2,
 \Phi_{02}=-\tau^2,\\
 E=-\frac{R}{12}-\rho \rd +\tau \td,
\end{eqnarray}
where $E$ is the real part of $\Psi_2=E+i H$,

c) the Bianchi equations (in which $\Psi_2, \Phi_{11}$ and $R$ are constants as a consequence of the homogeneity assumption):
\begin{eqnarray}
 \tho \Phi_{01} = -\rho (4 \Phi_{01}+5 \tau \rho), \label{bi1}\\
\thd \Phi_{01} = \rd \Phi_{01}-\rho \Phi_{12} +\tau(3 \overline{\Psi_2}+\tau \tbar-2 \Phi_{11}), \label{bi2} \\
\et \Phi_{01} = -\tau(4 \Phi_{01}+5 \tau \rho), \label{bi3} \\
\etd \Phi_{01}= -\tbar \Phi_{01} -\tau \Phi_{10} -\rho (3\overline{\Psi_2}+\rho \rd+2 \Phi_{11}), \label{bi4} \\
\tho(\mu^2) = 3 \rd(\overline{\Psi_2}-\Psi_2) -3 \tbar \Phi_{12}-3\tau \Phi_{21}, \label{bi5} \\
\et(\tbar^2) = 3\tbar(\overline{\Psi_2}-\Psi_2)+3 \rd \Phi_{10}-3 \rho \Phi_{21} . \label{bi6}
\end{eqnarray}

I will occasionally make use also of 0-weighted quantities $U,V$ (both real and with $U'=U$, $V'=V$) defined by 
\be
R=8(U-V)-16\rho \rd, \ 
\Phi_{11}=U+V-2 \rho \rd 
\ee

All these equations must be read as being accompanied by their `dashed' and complex conjugate versions.


\section{Classification of the homogeneous KS space-times}
In \cite{Jeffryes1984} KS space-times were classified, according to the behaviour of their
 GHP spin coefficients. The following cases were distinguished: class $I$ ($\rho \rho' \tau \tau' \neq 0$), class $I_N$ ($\rho \tau \tau' \neq 0$, $\rho'=0$), class $II$ 
($\tau \tau' \neq 0$, $\rho=\rho'=0$), class $III$ ($\rho \rho' \neq 0$, $\tau=\tau'=0$), class $III_N$ (
$\rho \neq 0$, $\rho'=\tau=\tau'=0$) and class $IV$ ($\rho=\rho'=\tau=\tau'=0$).
In classes $II$, $III$, $III_N$ and $IV$ conformal representants with regular Killing tensors always exist. In principle
one can use the results of \cite{Jeffryes1984} in order to obtain information about any possible homogeneous members in these classes.
This is however not obvious and therefore the analysis is presented from scratch in the unitary representant. 

All 0-weighted combinations of GHP quantities, such as $\rho \rd$ and $\tau \tbar$, are geometric invariants and therefore are constants in a 
homogeneous space-time: for each class the consequences of this elementary observation will be analysed separately.

\subsection{Class $I$}
Defining the variables $\phi, \phi'$ by
\begin{equation}\label{defPhi01}
  \td \Phi_{01}=-3 \rho \tau \td-2\rho \phi , \, \tau \Phi_{21}=-3 \rho' \tau \td-2\rho' \phi' ,
\end{equation}
homogeneity implies $\tho(\tau \td)=0$ and hence, using (\ref{thtau}), $\phi - \phibar =0$. If a class I conformal
representant would admit a regular
 Killing tensor, then~\cite{KS1_2} also $\phi + \phibar $ would be 0 and hence $\phi=0$. Then however Bianchi equation (\ref{bi1}), 
which in terms of the functions $\phi, \phi'$ reads
\[\tho \phi = \frac{2 \rho}{|\tau|^2}(|\tau|^4-|\phi|^2),\]
would lead to an inconsistency. If, on the other hand, a class I conformal representant would admit a semi-regular 
Killing tensor, then it would belong to the space-times discussed in \cite{KS1_2}. As the latter's unitary representant possesses at most 
a 1-dimensional isometry group, all homogeneous members of class $I$ must be given by the exceptional Killing tensor 
space-times discussed in \cite{Beke_et_al2011}. 
These were obtained by fixing the null tetrad $(\w{\omega}^a)$ such that $\rho = iQ r$, $\rd = -i r$ and $\tau =m$, after which the 
curvature components are given by
(\ref{defPhi00}) and
\be
R,\Psi_2,\Phi_{11},\Phi_{01} =
20(m^2+Q r^2), -\frac{8}{3}(m^2+Qr^2), -\frac{7}{2}(m^2-Qr^2),  -5\tau\rho ,\ \label{curv_caseI}
\ee
($Q=\pm 1$ and $r,m$ are real parameters). With respect to suitably chosen tetrads $(\w{\Omega}^a)$ or $(\w{\sigma}^a)$ 
the line-elements obtained were:\\

$Q=1$:
\begin{eqnarray}
\ud s^2 = 2({\w{\Omega}^1}^2+{\w{\Omega}^2}^2-Q {\w{\Omega}^3}^2+Q{\w{\Omega}^4}^2),\label{linelementOm} \\
 \w{\Omega}^1 = \frac{1}{2(m^2-r^2)}(2 m^2 \ud t - r \ud v) -\frac{r}{\mathcal{K}(x,y)}(y\ud x-x \ud y),\nonumber \\
 \w{\Omega}^2 = \frac{1}{\mathcal{K}(x,y)}(\sin v \ud x+\cos v \ud y ),\nonumber\\
 \w{\Omega}^3 = \frac{1}{\mathcal{K}(x,y)}(\cos v \ud x -\sin v \ud y) ,\nonumber\\
 \w{\Omega}^4 = \frac{m}{2(m^2-r^2)}(-2 r \ud t +\ud v) + \frac{m}{\mathcal{K}(x,y)}(y\ud x-x \ud y), \label{posQcaseA}.
 \end{eqnarray}
with
\be
\mathcal{K}(x,y)=1-\left(m^2-r^2\right)\left(x^2+y^2\right).
\ee

$Q=-1$:
\begin{eqnarray}
\ud s^2 = 2(\w{\sigma}^1\w{\sigma}^2-\w{\sigma}^3\w{\sigma}^4),\label{linelementSigma} \\
\w{\sigma}^1= \frac{1}{2\sqrt{2}}[(\mathcal{E}z^2-\frac{i}{m^2+r^2}\mathcal{E}^{-1})\ud x -\mathcal{E}\ud z],\nonumber\\
\w{\sigma}^3=\frac{1}{2(m^2+r^2)}[m \ud t+(m+r)z \ud x+r \ud y] ,\nonumber\\
\w{\sigma}^4=\frac{1}{2(m^2+r^2)}[m \ud t+(m-r)z \ud x-r \ud y],\label{negQ}
\end{eqnarray}
with
\be\label{Ecal}
\mathcal{E}=\exp(2\frac{r^2y+m^2t}{r^2+m^2}).
\ee

Note that the $m=r$ limit of (\ref{linelementSigma}) is conformally flat, while for $r\to m$  the metric (\ref{linelementOm}) 
has a singular limit of Petrov type D, in which the tetrad can be rewritten as
\begin{eqnarray}
 \w{\Omega}^1 = \ud(v-u)+2 m x \,\ud y ,\nonumber\\
 \w{\Omega}^2 = \sin 2mv \,\ud x+\cos 2mv \,\ud y,\nonumber\\
 \w{\Omega}^3 = \cos 2mv \,\ud x-\sin 2mv \,\ud y,\nonumber\\
 \w{\Omega}^4 =  \ud u -2 m x \,\ud y.\label{m=r metric}
 \label{posQcaseB}
\end{eqnarray}

In all cases the dimension of the isometry group is 4 (there is no isotropy as $\rho$ and $\tau$ are $\neq 0$).

\subsection{Class $I_N$}

When $\rho' = 0$ (\ref{etrho}) implies $\Phi_{12}=0$. There are then always\cite{McLenVdB1993} conformal representants admitting a regular or semi-regular 
Killing tensor. As in the previous paragraph homogeneity implies that $\phi$ is real and the same argument as before shows that the regular case 
leads to an inconsistency. The semi-regular case was treated in \cite{McLenVdB1993}: there however only the hypersurface 
homogeneous situation was considered and the possibility was overlooked (see equations (2.15)) that all the Cartan invariants could 
be space-time constants. The Bianchi equations, together with the assumption that $\phi,\Psi_2,U$ and $V$ are constant,
 imply then $\phi=\pm |\tau|^2$, $U=-|\tau|^2/2$, $V=2 \phi-|\tau|^2$ and 
\[
 \Psi_2=\frac{4}{3}|\tau|^{-2}(\phi+2 |\tau|^2)(\phi-|\tau|^2).
\]
Writing $|\tau|=m$ (constant) the non-conformally flat cases are characterised by 
\[
 R,\Psi_2,\Phi_{11},\Phi_{01}=20m^2, -\frac{8}{3}m^2,-\frac{7}{2}m^2,-5 i m^2,
\]
which, when compared with (\ref{curv_caseI}), shows that the corresponding metrics should be obtainable by taking the $Q\to \infty$ limit 
of class I (after replacing $r$ by $r/Q$). As the class I metrics (\ref{linelementOm},\ref{linelementSigma},\ref{posQcaseB}) were constructed 
assuming explicitly $Q=\pm 1$, this limit is not easy to recognise and therefore these metrics are constructed below from scratch. 
First the tetrad is fixed by means of a boost and a rotation such that 
$\rho = i m$ and $\tau=\pi=m$. The Cartan equations become then
\begin{eqnarray}
\ud \w{\omega}^1 = -2 m \w{\omega}^4 \wedge (i\w{\omega}^1+\w{\omega}^3),\label{CartanINa} \\
\ud \w{\omega}^3 = 2 i m \w{\omega}^2 \wedge \w{\omega}^1 -2 m \w{\omega}^3\wedge( \w{\omega}^1+\w{\omega}^2),\label{CartanINb} \\
\ud \w{\omega}^4 = 2 m \w{\omega}^4 \wedge ( \w{\omega}^1+\w{\omega}^2).\label{CartanINc}
\end{eqnarray}
It follows that $i ( \w{\omega}^1-\w{\omega}^2)-\w{\omega}^4$ is closed and that (the dual vector field of) $\w{\omega}^4$ is
 hypersurface-orthogonal. This allows one to write $\w{\omega}^4=x \ud u$ ($\ud x \wedge \ud u \neq 0$), after which (\ref{CartanINc}) implies
\[
\w{\omega}^1+\w{\omega}^2=-\frac{1}{2 m x} \ud x + v \ud u \ \ (\ud x \wedge \ud u \wedge \ud v \neq 0).
\]
 Writing $i ( \w{\omega}^1-\w{\omega}^2)=\w{\omega}^4+ 2\ud y$, ($\overline{\ref{CartanINa}}$) shows
that \[\w{\omega}^3=\frac{1}{4 m x}\ud v -\ud y+f \ud u,\] with $f=-(x^2+v^2)/(4 x)+F$ and $F$ an arbitrary function of $u$. 
As the resulting spin coefficients and curvature components depend on the parameter $m$ only, it follows\cite{Karlhede} that a coordinate transformation 
\emph{must} exist making $F=0$\footnote{the explicit construction of this transformation 
is not obvious}. The null tetrad becomes herewith
\begin{eqnarray}
  \w{\omega}^3 &= \frac{1}{4mx}\ud v -\ud y -\frac{1}{4x}(x^2+v^2)\ud u,\ 
 \w{\omega}^4 = 2 x \ud y \label{tetrad_IN} \\
 \w{\omega}^1 &=  -\frac{1}{4mx}\ud x +\frac{1}{2}(v-ix)\ud u -i\ud y .
\end{eqnarray}
After a coordinate transformation $v\to v/(mx)$, the line-element can then be written as
\be\label{ds_metricIN}
 \ud s^2 = \frac{1}{8m^2 x^2}(\ud x^2+8(v^2-m^2x^4)\ud u^2)+2(\ud y+x \ud u)^2-\frac{1}{2m^2 x}\ud u \ud v .\label{compact_IN}
\ee
Again the maximal dimension of the isometry group is 4.

\subsection{Class $II$}
Imposing $\rho=\rd=0$ (and hence $\Phi_{00}=\Phi_{22}=0$), the Bianchi equations, together with the constancy of 
$R,\Phi_{11},\Psi_2$ and $|\tau|=m$, 
lead to
 $\Phi_{01}=\Phi_{12}=0$ and
\[
 R,\Psi_2,\Phi_{11}=-4m^2-8V,\frac{2}{3}(V-m^2),V-\frac{1}{2}m^2.
\]
The null tetrad can now be partially fixed by rotating such that $\tau=m$. Switching to the Newman-Penrose formalism, it furthermore follows
from the Bianchi equations
that the spin coefficients $\epsilon$ and $\gamma$ are real and that $\beta=\overline{\alpha}$. The Cartan equations
 show then that 
$\w{\omega}^1-\w{\omega}^2$ is closed and that (the
dual vector fields of) $\w{\omega}^3$ and $\w{\omega}^4$ are hypersurface-orthogonal. This allows one to partially fix a boost such that 
$\w{\omega}^3$ is closed too, implying $\epsilon=0$ and $\alpha=\tau/2$. The Newman-Penrose equations reduce then to
\be
 D\gamma=2 m^2+2 V ,\ \delta \gamma=\overline{\delta}\gamma = 0, \label{caseII_gamma_eq}
\ee
while the Cartan equations become
\begin{eqnarray}
 \ud \w{\omega}^1 = 2 \tau \w{\omega}^3\wedge \w{\omega}^4,\label{Cartan_IIa}\\
\ud \w{\omega}^3=0,\label{Cartan_IIb}\\
\ud \w{\omega}^4=2\gamma \w{\omega}^4\wedge\w{\omega}^3.\label{Cartan_IIc}
\end{eqnarray}
It follows that $\w{\omega}^3=\ud u$ and $\w{\omega}^4=e^p \ud v$, with $p$ a function of $u$ and $v$: by (\ref{caseII_gamma_eq}) 
$2\w{\omega}^3\w{\omega}^4$ is then the metric of a two-space of constant curvature. One should now distinguish the 
flat and non-flat cases:\\

a) when $V+m^2\neq 0$ the coordinate $v$ can be re-defined such that $\w{\omega}^4=kv^{-2}(\ud v-\ud u)$,
with, by (\ref{caseII_gamma_eq}), $2k=-1/(m^2+V)$. 
One has then $\gamma=1/v$ and, putting $\w{\omega}^1-\w{\omega}^2=i \ud y$ (\ref{Cartan_IIa}) can be integrated to give
\begin{eqnarray}
\w{\omega}^1+\w{\omega}^2=\ud x+\frac{\ell}{v}(\ud v -\ud u), \ \w{\omega}^1-\w{\omega}^2=i \ud y, \nonumber\\
\w{\omega}^3=\ud u,\ \w{\omega}^4=kv^{-2}(\ud v-\ud u),\label{classII_metric}
\end{eqnarray}
with $\ell=2m/(m^2+V)$. Note that the case $V=m^2$ is conformally flat. \\

b) when $V+m^2=0$ the integration is straightforward and one obtains
\be
\w{\omega}^1=2 m(\ud x+i \ud y +u\ud v),\w{\omega}^3=\ud u,\ \w{\omega}^4=\ud v,\label{classII_metric_spec}
\ee
yielding the line element
\be
 \ud s^2 = 4 m^2((\ud x+u\ud v)^2+\ud y^2)-\ud u \ud v.
\ee
As there is a residual 1-dimensional group of boost isotropies, both families (a) and (b) admit a 5 dimensional isometry group.

\subsection{Class $III$}
This class is the Sachs' transform~\cite{GHP} of class $II$. All calculations are therefore similar to those of the 
previous paragraph, but now 
a distinction has to be made between the cases $\w{\omega}^3\pm\w{\omega}^4$ being closed: this is responsible for the appearance of 
the extra parameter $Q=\pm 1$ below. Starting with $\tau=\td=0$ and
the constancy of $R, \Phi_{11}, \Psi_2$ and $\rho\rd$, the only other non-0 curvature components turn out to be $\Phi_{00}=-\rho^2$ and 
$\Phi_{22}=-\rd^2$. Fixing a boost such that $\rho = i Q r$, $\mu=i r$ ($r\in \mathbb{R}$ constant), the Newman-Penrose and Bianchi equations imply
\[
 R,\Psi_2,\Phi_{11}=8(U-\frac{1}{2} Qr^2),-\frac{2}{3}(U+Q r^2),U+\frac{1}{2}Qr^2,
\]
with the spin coefficients satisfying $\overline{\epsilon}=-\epsilon,\overline{\gamma}=-\gamma,\beta=-\overline{\alpha}$. 
The Cartan equations become
\begin{eqnarray}
\ud \w{\omega}^1 = \w{\omega}^1\wedge(2 \alpha \w{\omega}^2  + (2 \gamma - ir) \w{\omega}^3 +(2\epsilon-iQr)\w{\omega}^4),\label{Cartan_IIIa}\\
 \ud \w{\omega}^3 = 2 i Q r \w{\omega}^2 \wedge \w{\omega}^1,\label{Cartan_IIIb}\\
\ud \w{\omega}^4 = 2 i r \w{\omega}^2 \wedge \w{\omega}^1. \label{Cartan_IIIc}
\end{eqnarray}

As (the dual vector field of) $\w{\omega}^1$ is  hypersurface-orthogonal, one can partially fix a spatial rotation such that
 $\w{\omega}^1=P^{-1} \ud \zeta$ with $P$ real and $\zeta$ complex. 
As $\w{\omega}^4-Q\w{\omega}^3$ is closed, a coordinate $u$ is defined locally by $2 \ud u =\w{\omega}^4-Q\w{\omega}^3$. Introducing a fourth
coordinate $w$ such that $\w{\omega}^3$ is spanned by $\ud \zeta, \ud \overline{\zeta}, \ud w$, equation (\ref{Cartan_IIIa}) and its complex 
conjugate imply that 
$P=P(\zeta, \overline{\zeta})$, $\epsilon=iQr/2$, $\gamma = ir/2$, $\beta=-\overline{\alpha}$ and $\alpha=\alpha(\zeta, \overline{\zeta})$.
Herewith the remaining Newman-Penrose equations are, in analogy with (\ref{caseII_gamma_eq}), given by
\be
\delta \alpha+\overline{\delta \alpha}-4\alpha \overline{\alpha} = 2U-2Qr^2,\ D\alpha=\Delta \alpha = 0,
\ee
expressing that $2\w{\omega}^1\w{\omega}^2=P^{-2}\ud \zeta \ud \overline{\zeta}$ is the metric of a two-space of constant curvature. 
Defining standard coordinates in this two-space by 
\[
 \w{\omega}^1=(1+\frac{k}{4}x^2)^{-1}(\ud x + i\ud y)
\]
and writing $\w{\omega}^3=f\ud x +g\ud y+h \ud w$, equation (\ref{Cartan_IIIb}) shows that $f,g,h$ are functions of $x,y$ and $w$, allowing one 
to put $w=1$ and $g=0$. Integration of (\ref{Cartan_IIIb}) gives then 
\be\label{f_expression}
 f=4Q r \frac{xy}{(1+\frac{k}{4}x^2)^2},
\ee
after which the line-element becomes ($w=v-Q u$)
\be\label{classIIImetric}
\ud s^2 = 2 (1+\frac{k}{4}x^2)^{-2} (\ud x^2+x^2 \ud y^2) -2 Q [(f\ud x+\ud v)^2-\ud u^2].
\ee
The constant $k$ is related to $U$ and $r$ by
\be
 U=Q r^2+\frac{k}{8}.
\ee
There is an obvious residual isotropy group of spatial rotations and hence these space-times admit a 5-dimensional isometry group.
The G\"odel metric is obtained as the special case $U=0,Q=1$. Note that the case $k+16 Q r^2=0$ is excluded, as it gives rise to a conformally flat metric.

\subsection{Class $III_N$}
This is the special case $\rd=0$ of class $III$: fixing the boost such that  $\rho = i r$ ($r\in \mathbb{R}$ constant) and 
proceeding in exactly the same way as in the previous 
paragraph, one recovers the expressions for $\w{\omega}^1, \w{\omega^3}$ (but with $Q$ replaced by 1), while $\w{\omega}^4$ becomes exact.
The line-element reads then
\be\label{classIIINmetric}
\ud s^2 = 2 (1+\frac{k}{4}x^2)^{-2} (\ud x^2+x^2 \ud y^2) -2 \ud u \ud v -2 f \ud x \ud u,
\ee
with 
\be\label{IIINf_expression}
 f=4Q r \frac{xy}{(1+\frac{k}{4}x^2)^2}
\ee
and $ U=\frac{k}{8}$ (now the flat case $k=0$ is excluded). As before there is a 5-dimensional isometry group.

\subsection{Class $IV$}

Here $\rho=\rd=\tau=\td=0$ and the only non-vanishing curvature components are the constants
\[
 R,\Psi_2,\Phi_{11}= 8(U-V), \frac{2}{3}(V-U), U+V.
\]
The four basis vector fields are hypersurface-orthogonal and one can partially fix a boost and a spatial rotation such that
\be
\w{\omega}^1 = P^{-1} \ud \zeta,\ 
\w{\omega}^3 = Q^{-1}\ud u,\ 
\w{\omega}^4 = Q^{-1}\ud v, \label{classIV}
\ee
with $P$ and $Q$ depending on $\zeta, \overline{\zeta}$ and $u,v$ respectively. The surviving Newman-Penrose equations are, with real
$\epsilon,\gamma$:
\begin{eqnarray}
D\gamma-\Delta \epsilon +4 \epsilon \gamma = 2 V ,\ \delta \gamma=\delta \epsilon = 0,\label{IVnpa}\\
 \delta \alpha+\overline{\delta} \overline{\alpha}-4\alpha \overline{\alpha} = 2U,\ D\alpha=\Delta \alpha = 0,\label{IVnpb}
\end{eqnarray}
expressing that these space-time are products of two constant-curvature two-spaces, of signature 0 and 2 respectively. Alternatively they can be characterised as the conformally
symmetric\cite{ChakiGupta,Kramer} space-times of Petrov type D. With $Q=e^q$ and $P=e^p$ one obtains
\begin{eqnarray}
p_{\zeta \overline{\zeta}}= 2 U e^{-2p},\\
q_{uv}= 2 V e^{-2 q},
\end{eqnarray}
i.e.~the Ricci scalars of the two-spaces $(u=const,v=const)$ and $(\zeta=const)$ are respectively given by $8 U$ and $8 V$. The line-element can be written as
\begin{equation}
  \ud s^2= 2 (1+2U \zeta \overline{\zeta})^{-2} \ud \zeta \ud \overline{\zeta} -2 (1+2 Vu v)^{-2} \ud u \ud v ,\label{IVds2}
\end{equation}
with $U\neq V$ as otherwise the space-time is conformally flat.
Because of the residual freedom of boosts and spatial rotations, (the non-conformally flat members of) these space-times admit a 6-dimensional isometry group.

\section{Interpretation}

With the exception of (\ref{classIIImetric}), which for $U=0,Q=1$ reduces to the G\"odel metric and of (\ref{IVds2}), which for $U+V>0$ is the 
$\Lambda \neq0$ generalisation of the 
Bertotti-Robinson metric\cite{Bertotti,Robinson} (the unique conformally flat
non-null Einstein-Maxwell solution when $\Lambda=0$), none of the above Killing spinor space-times has an immediate physical interpretation. 
Some of their conformal representants however might be interpreted as Einstein-Maxwell, perfect fluid or pure radiation space-times. 
Addressing this question directly in coordinates, leads to insuperable problems. 
A better strategy is to stay within the Newman-Penrose formalism and to express the spin coefficients and curvature 
components of the conformally transformed metric in terms of those of the unitary representant and of the directional derivatives $\partial_a \Omega$ of the conformal factor $\Omega$. This still leads to a complicated algebraic and differential consistency analysis 
for the unknowns $D\Omega, \Delta \Omega, \delta \Omega, \overline{\delta} \Omega$ and $n_1, n_2, n_3, n_4$ (in the 
case of a perfect fluid $n_a =\sqrt{w+p} u_a$ with $u^a$ the 4-velocity), or 
(in the case of a Maxwell field) the 6 components $\Phi_0, \Phi_1, \Phi_2$ of the Maxwell spinor. 
In the following paragraphs this procedure (for details of which the reader is referred to \cite{Beke_et_al2011}) is applied to the different classes of 
homogeneous KS space-times discussed above.

\subsection{Classes $I, I_N, II$ and $III_N$}
For class $I$, determined by (\ref{linelementOm}-\ref{posQcaseA}), 
(\ref{linelementSigma}-\ref{negQ}), (\ref{posQcaseB}),
it was shown\cite{Beke_et_al2011} that there are no perfect 
fluid (neither aligned nor non-aligned), pure radiation nor Einstein-Maxwell (neither singly nor doubly aligned) interpretations. 
The same conclusions hold for classes $I_N$ and $III_N$, determined by (\ref{tetrad_IN}) and (\ref{classIIINmetric}) respectively:  the analysis is quasi-identical, although some care 
is necessary to exclude the singly aligned Einstein-Maxwell case, as the vanishing of the Maxwell spinor components $\Phi_0$ or $\Phi_2$ is now not equivalent).

For class $II$, determined by (\ref{classII_metric}) and (\ref{classII_metric_spec}), the Einstein-Maxwell case demands some more work than in 
$I$ or $I_N$, as $D \Omega =0$ cannot be 
excluded a priori on the basis of equations (69a) or (69f) of \cite{Beke_et_al2011}. As all results ---for perfect fluids, pure radiation and Maxwell fields--- are again negative (and identical to those 
for class I) and as the calculations are quite long and little 
illuminating, they will not be repeated here.

\subsection{Class $III$}
The class $III$ metrics, determined by (\ref{classIIImetric}), clearly exhibit local rotational symmetry (LRS)\cite{StewartEllis}: they belong 
to the LRS I family when $Q=1$ 
and to the LRS III family when $Q=-1$. A search for a perfect fluid interpretation along the lines of \cite{Beke_et_al2011} reveals that there 
is no such interpretation when $Q=-1$: this is in agreement with the fact that the class III metrics all have a purely electric Weyl tensor
($\Psi_2=-(k+16Qr^2)/12)$ and that no LRS III purely electric perfect fluids exist\cite{Lodethesis}. On the other hand, 
when $Q=1$ a perfect fluid interpretation does exist and is necessarily aligned: multiplying the metrics (\ref{classIIImetric}) with a scale factor
 $\Omega=1/\sinh((k/2+4 r^2)^{1/2} u)$ transforms (\ref{classIIImetric}) into a purely
electric ``stiff fluid'' metric ($\ud p / \ud w =1$). Purely electric LRS I metrics are generalisations of the G\"odel-metric: they are stiff
fluids, the metric of which can be written in the standard form\cite{Lodethesis},
\begin{equation}\label{PEPFLRSI}
 \ud s^2 = \frac{1}{j(x)^2} \left[ \frac{\ud x^2}{j(x)^2} -(\ud t +q r^2 e^{\psi} \ud \phi)^2+e^{2\psi}(\ud r^2+r^2\ud \phi^2),\right]
\end{equation}
with $e^{-\psi} =1+\frac{k}{4}r^2$ and $j^2=(\frac{k}{2}+q^2)x^2+c_1 x+ c_2$. In fact, dropping the factor $j^{-2}$ from (\ref{PEPFLRSI}), 
a coordinate transformation brings these metrics in the form (\ref{classIIImetric}). This shows that for LRS I perfect fluids the condition of
 being purely electric is equivalent with being conformally homogeneous. Again, as for class $I$ there are no (doubly nor singly) aligned 
Einstein-Maxwell nor pure radiation interpretations.

\subsection{Class $IV$}
As remarked at the end of the previous section the class $IV$ metric (\ref{classIV}) itself represents an aligned non-null Einstein-Maxwell field (provided $U+V>0$). Again one can 
show that there is no pure radiation interpretation
(aligned nor non-aligned), but both aligned and non-aligned perfect fluid interpretations do exist: surprisingly the general solution for the 
non-aligned case can be given in explicit form. For a detailed discussion of the ensuing solutions, see \cite{Confdecomp_CQG}

\section{Discussion}
Homogeneous KS space-times, being of Petrov type D, admit either a 4, 5, or 6-dimensional group $G_n$ of isometries. The following list of
space-times exhausts these three families:\\

\noindent
$G_4$:
class $I$, determined by (\ref{linelementOm}-\ref{posQcaseA}), 
(\ref{linelementSigma}-\ref{negQ}), (\ref{posQcaseB})
and class $I_N$, determined by (\ref{tetrad_IN})\\

\noindent
$G_5$:
class $II$, determined by (\ref{classII_metric}) and (\ref{classII_metric_spec}),
class $III$, determined by (\ref{classIIImetric}) and
class $III_N$, determined by (\ref{classIIINmetric})\\

\noindent
$G_6$:
class $IV$, determined by (\ref{IVds2}).\\

None of these metrics admits (a conformal representant with) a pure radiation interpretation. A perfect fluid interpretation only exists for 
the class III
metric (\ref{classIIImetric}) with $Q=1$ and the class IV metrics. The $Q=-$ class III metrics are precisely the purely electric LRS I 
stiff fluid metrics, generalising the G\"odel-metric, while
the class IV metrics are either aligned and form a sub-class of the LRS II perfect fluids, or they are non-aligned and then can be given in 
explicit form. The resulting
space-times, discussed in \cite{Confdecomp_CQG}, in general admit no symmetries. 

A different problem is related to the non-nul and non-aligned Petrov type D Einstein-Maxwell solutions. Very few of these being explicitly known (Griffiths' 1986 solution\cite{Griffiths}
 being the only one known to the author), it remains a tantalising question whether any such solution might be constructable by conformally transforming any of 
the KS space-times discovered to date.

\noindent

\section*{References}


\providecommand{\newblock}{}

\end{document}